\documentclass[aps,prb,twocolumn,superscriptaddress, showpacs]{revtex4}
\usepackage{graphicx} 

\begin{document}
 

\title{Universal magnetic and structural behaviors in the iron arsenides}



\author{Stephen D. Wilson}
\affiliation{Materials Science Division, Lawrence Berkeley National Laboratory, Berkeley, California 94720}
\author{C. R. Rotundu}%
\affiliation{Materials Science Division, Lawrence Berkeley National Laboratory, Berkeley, California 94720}
\author{Z. Yamani}
\affiliation{Canadian Neutron Beam Centre, National Research Council, Chalk River Laboratories, Chalk River, Ontario, K0J 1P0, Canada}%
\author{P. N. Valdivia}
\affiliation{Materials Science and Engineering Department, University of California, Berkeley, California 94720}%
\author{B. Freelon}
\affiliation{Physics Department, University of California, Berkeley, California 94720}%
\author{E. Bourret-Courchesne}%
\affiliation{Materials Science Division, Lawrence Berkeley National Laboratory, Berkeley, California 94720}
\author{R. J. Birgeneau}
\affiliation{Materials Science Division, Lawrence Berkeley National Laboratory, Berkeley, California 94720}
\affiliation{Materials Science and Engineering Department, University of California, Berkeley, California 94720}%
\affiliation{Physics Department, University of California, Berkeley, California 94720}%


\date{\today}

\begin{abstract}
Commonalities among the order parameters of the ubiquitous antiferromagnetism present in the parent compounds of the iron arsenide high temperature superconductors are explored.  Additionally, comparison is made between the well established two-dimensional Heisenberg-Ising magnet, K$_2$NiF$_4$ and iron arsenide systems residing at a critical point whose structural and magnetic phase transitions coincide.  In particular, analysis is presented regarding two distinct classes of phase transition behavior reflected in the development of antiferromagnetic and structural order in the three main classes of iron arsenide superconductors.  Two distinct universality classes are mirrored in their magnetic phase transitions which empirically are determined by the proximity of the coupled structural and magnetic phase transitions in these materials.  
\end{abstract}

\pacs{74.70.Dd; 75.25.+z; 75.50.Ee; 75.40.Cx}

\maketitle

\section{Introduction}
Understanding the fundamental properties of the underlying magnetic and structural order in the newly discovered iron-based high temperature superconductors (high-T$_c$) is an essential step in the eventual resolution of magnetism's role in the superconducting pairing in these systems.  The key unifying feature between the iron arsenide class of high-T$_c$ superconductors and the well known cuprate high-T$_c$s is the universal presence of an antiferromagnetically (AF) ordered state in close proximity to the development of superconductivity within their respective phase diagrams\cite{kastner, huangLaFeAsO, chu}.  The undoped, parent systems of both classes of high-T$_c$s exhibit long-range, AF order\cite{kastner, cruz} that is suppressed upon doping and either vanishes\cite{luetkens} or weakly competes\cite{zhaoce, mcqueeneyBa122Co} with the onset of superconductivity.  While previous exploration of critical properties intrinsic to the static AF order in the cuprates led to fundamental insights regarding the interactions governing the spin behavior in those systems\cite{birgeneauLa2CuO4}, experiments probing the detailed behavior of magnetism in the iron pnictide systems are only just beginning. 

The ordered spin structures of the LaFeAsO(1111) \cite{cruz}, NaFeAs(111) \cite{shiliangNaFeAs}, and BaFe$_2$As$_2$(122)-type\cite{huangBa122} iron arsenide compounds possess a common unidirectional antiferromagnetic ordering as shown in Fig. 1 where the AF propagation vector points both along the long-axis within the basal plane and along the out-of-plane, $c$-axis.   The in-plane component of the AF wave vector is determined by an accompanying or preceding structural distortion from tetragonal to orthorhombic symmetry in which the $a$-axis within the iron layers is slightly elongated.  While the exchange couplings and spin dynamics intrinsic to this magnetic structure in the new class of iron-based high-T$_c$s have been a topic of considerable focus, relatively little attention has been given to any precise investigation of the magnetic phase transition behavior in these systems.  Indeed, within the bilayer AEFe$_2$As$_2$ class of pnictides (where single-crystals are most easily synthesized) a number of early studies suggested first-order behavior in the onset of the AF phase\cite{kofu, krellner, goldman, kittigawa}. While these reports seemingly precluded any subsequent study of critical behavior in these systems, separate reports also showed continuous, second-order magnetic phase transitions in a number of the same materials\cite{tegel, rotter, matan}. Recently, new single-crystal studies have appeared showing continuous phase transitions in a number of iron pnictide variants\cite{wilson, mcqueeneyBa122Co, lelandBa122Ni, haydenBa122Co} thereby suggesting that improvements in sample quality and diversity may now facilitate a reliable comparison between systems.

\begin{figure}
		\includegraphics[scale=0.5]{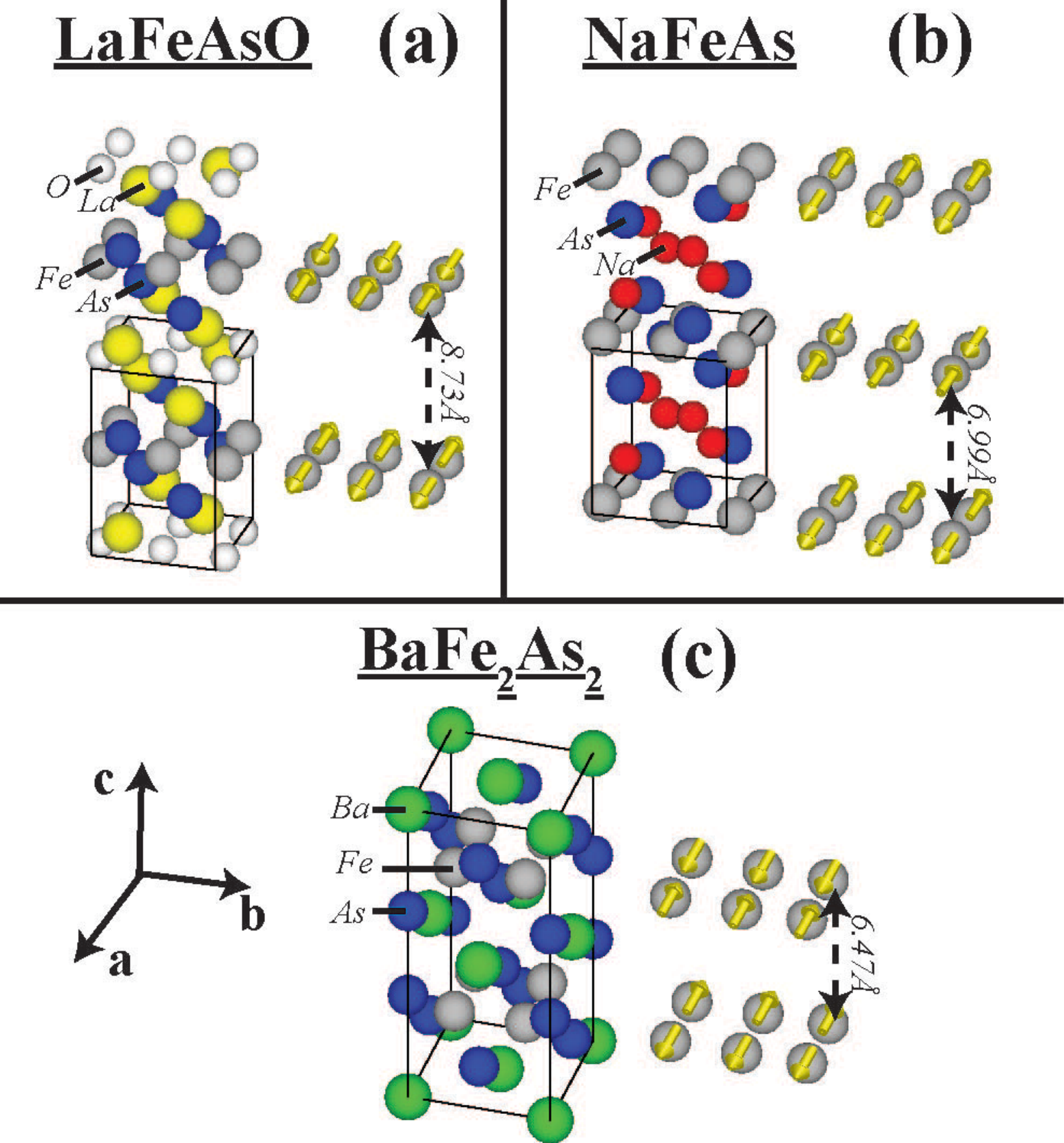}
		\caption{Crystal structures of representative materials from the (a) REFeAsO, (b) AMFeAs, and (c) AFe$_2$As$_2$ classes of iron arsenides.  Boxes show the enclosed chemical unit cells of each system.  To the right of each chemical structure are plotted the ordered spin systems.  Moment directions are plotted as yellow arrows located on the Fe-sites.}
	\label{fig:Fig1}
\end{figure}

\begin{figure}
		\includegraphics[scale=0.5]{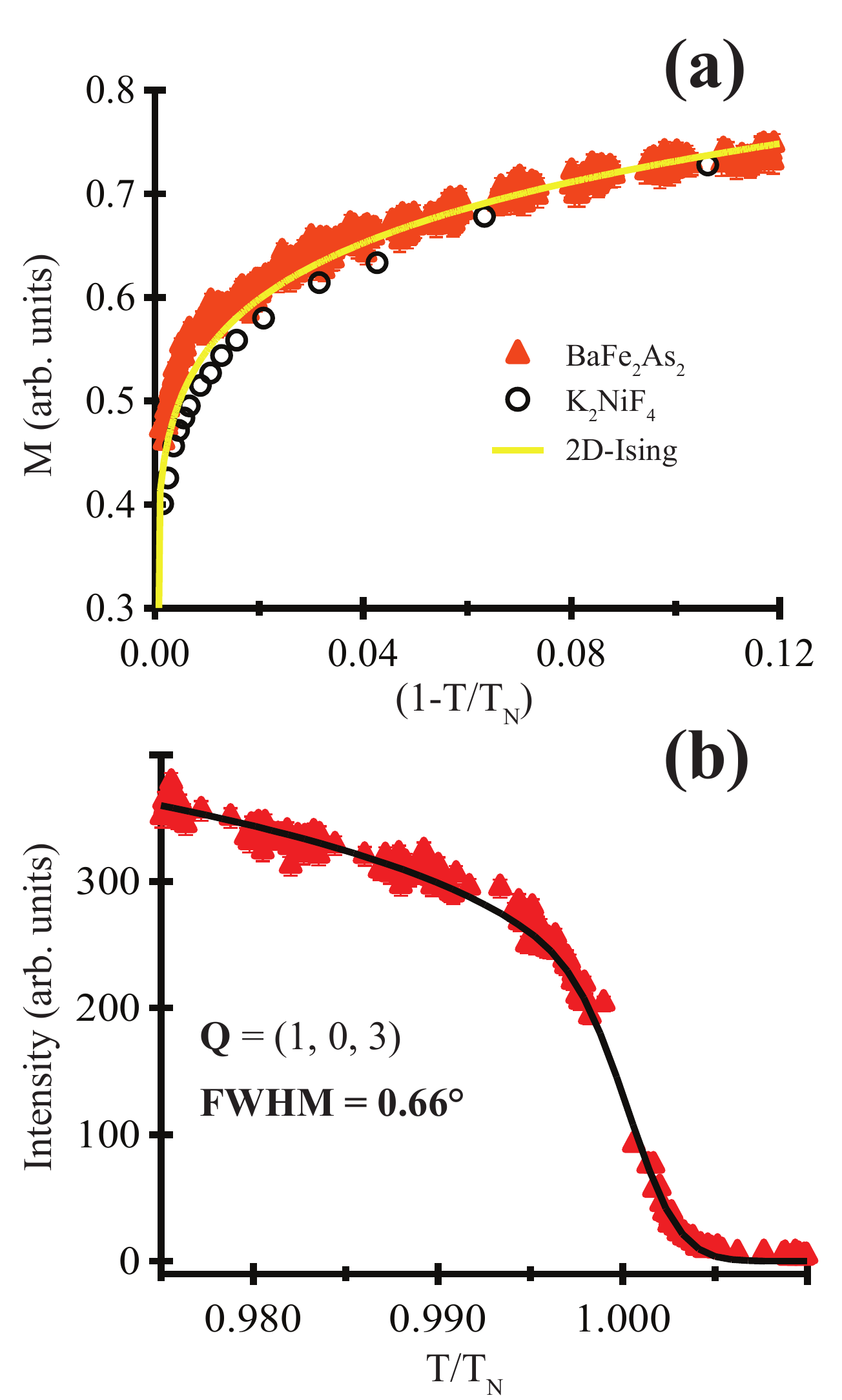}
		\caption{Comparison plots of the magnetic order parameters in BaFe$_2$As$_2$ and K$_2$NiF$_4$ with normalization described in the text.  The solid yellow line denotes the expected behavior for the ideal 2D-Ising system.  (a)Neutron measurements of M($\frac{T}{T_N}$) for BaFe$_2$As$_2$ and K$_2$NiF$_4$ taken from Refs. 18 and 21 respectively.  (b) Plot showing M$^2$($\frac{T}{T_N}$) for BaFe$_2$As$_2$ fit with a power law model convolved with a Gaussian distribution of T$_N$s as described in the text.}
	\label{fig:Fig2}
\end{figure}
In this article, we present our analysis of the magnetic and structural order parameters in the three main classes of iron pnictide superconductors:  AFe$_2$As$_2$ (A-site = Ba, Sr, Eu, ...), AMFeAs (AM = alkali metal = Na, Li), and REFeAsO (RE = rare earth = La, Pr, Ce, ...).  Undoped systems within the bilayer, 122 class exhibit a two-dimensional (2D) Ising-like magnetism whose order parameters are shown to mirror closely the spin behavior in the prototypical 2D-Heisenberg-Ising system K$_2$NiF$_4$.  Once the 122 systems are perturbed and the structural and magnetic phase transitions no longer coincide in temperature, a fundamentally different magnetic behavior emerges.  In this regime, when T$_N<$T$_S$, a dramatically altered development of the magnetic order parameter appears and collapses onto a seemingly universal curve with a critical exponent of $\beta \approx 0.25$.  This reflects a crossover from the coincident T$_N=$T$_S$, 2D-Ising-like behavior to a distinctly separate universality class where the coupling between T$_S$ and T$_N$ has been substantially weakened.  Whereas the previously observed coupling between the primary structural and magnetic order parameters in the undoped 122 systems renders identical critical exponents for both phase transitions, surprisingly, we find that far from the critical point (where T$_N<$T$_S$) both critical exponents modeling the structural and magnetic phase behavior in the iron arsenides change in an identical fashion.  Thus even for systems such as LaFeAsO, where T$_s$=1.1T$_N$, the structural, orthorhombic distortion develops in a manner identical to the onset of AF order in the material.   These empirical observations of potentially universal magnetic and structural behavior among the various classes of iron pnictide high-T$_c$s hold promising implications for the identification of a common role for magnetism within their superconducting properties.

\section{Magnetism in K$_2$N\MakeLowercase{i}F$_4$ and B\MakeLowercase{a}F\MakeLowercase{e}$_2$A\MakeLowercase{s}$_2$}  
Our previous investigation of AF order in the BaFe$_2$As$_2$ (Ba-122) compound\cite{wilson} revealed that the magnetic order parameter was well modeled by a simple power law behavior of $(1-\frac{T}{T_N})^{\beta}$ with an exponent of $\beta=0.125$ thereby suggesting that the universality class of the AF phase transition was the same as that of the 2D-Ising model.  While a weakly first order component to the phase transition could not be entirely precluded, our previous analysis placed a limit on any possible linear, first order trade-off between phases to less than $0.5$K of T$_N$.  The remarkable range in reduced temperature over which the simple power law models the magnetic phase transition in BaFe$_{2}$As$_{2}$ is reminiscent of phase behavior in known low dimensional magnets such as the prototypical 2D magnet K$_{2}$NiF$_{4} $\cite{birgeneauK2Nif4} which stands as a useful standard for comparison when examining the phase transition behavior in BaFe$_{2}$As$_{2}$.

K$_2$NiF$_4$ is a model 2D system with a crossover from 2D Heisenberg behavior at low temperatures to anisotropy-driven 2D Ising behavior in the vicinity of T$_N$.  Figure 2 (a) displays a comparison of the magnetic order parameters, $M$, as a function of $(1-T/T_N)$ for both systems K$_{2}$NiF$_{4}$ and BaFe$_{2}$As$_{2}$ where data for BaFe$_{2}$As$_{2}$ are taken from Ref. 18  and the data for K$_{2}$NiF$_{4}$ are extracted from Ref. 21.  In order to facilitate a comparison of the phase transition behavior in each respective system, all data were normalized at $(1-(T/T_N)) = 0.1$, where $\approx 75\%$ of the saturated moment had been reached.  Examination of this comparison highlights the striking similarity between the evolution of the magnetic order parameters in Ba-122 and K$_{2}$NiF$_{4}$ where the critical behaviors of both systems with freely refined exponents of $\beta = 0.10$ and $\beta = 0.14$ respectively bracket the ideal 2D Ising curve with $\beta=0.125$. In particular, this demonstrates explicitly that the sharp onset of antiferromagnetism in BaFe$_2$As$_2$ mirrors that of a known second order 2D-Heisenberg-Ising AF phase transition and thus that it cannot be easily dismissed as an artifact from an obscured first order phase transition.

At temperatures immediately above $T_N$, appreciably more magnetic scattering at the 3D ordering wave vector persists in BaFe$_{2}$As$_{2}$ relative to K$_{2}$NiF$_{4}$.  The origin of this tail of scattering above $T_N$ is likely due to a small distribution of AF ordering temperatures throughout the bulk of the sample.  This can be confirmed by fitting the expected power law behavior of the order parameter weighted by a Gaussian distribution of $T_N$'s within the sample.  The result of such a fit to the form $M(T)^2 = B^2\int(1-\frac{T}{T_N})^{2\beta}e^{-(T-\left\langle T_N\right\rangle)/2\sigma^2}$ is shown in Fig. 2 (b), and the fit gives values of $\langle T\rangle = 136.12 \pm 0.03$K, $\beta = 0.106 \pm 0.002$, and $\sigma = 0.28 \pm 0.02$K.  The power law was forced to 0 for $T>T_N$, and only data in the range $0.85$K$<\frac{T}{T_N}<1.025$K were included.  The excellent agreement between this Gaussian-broadened power law description and the magnetic order parameter in BaFe$_2$As$_2$ clearly demonstrates that the phase transition in its entirety can be effectively modeled by a simple power law with a $\beta\approx 0.125$ and a slightly broadened $T_N$ with full width at half-maximum, FWHM $= 0.0048$ in reduced temperature.  Additionally, our subsequent fitting of the same data to the first order form $M^2(T) = (A_s+B(1-\frac{T}{T_N})^\beta)^{2}$ yielded negligible $A_s$ values and identical $\beta$ values to the previous $A_s \equiv 0$ fits, thus further supporting our claim of continuous behavior in BaFe$_2$As$_2$.

\begin{figure}
\includegraphics[scale=0.5]{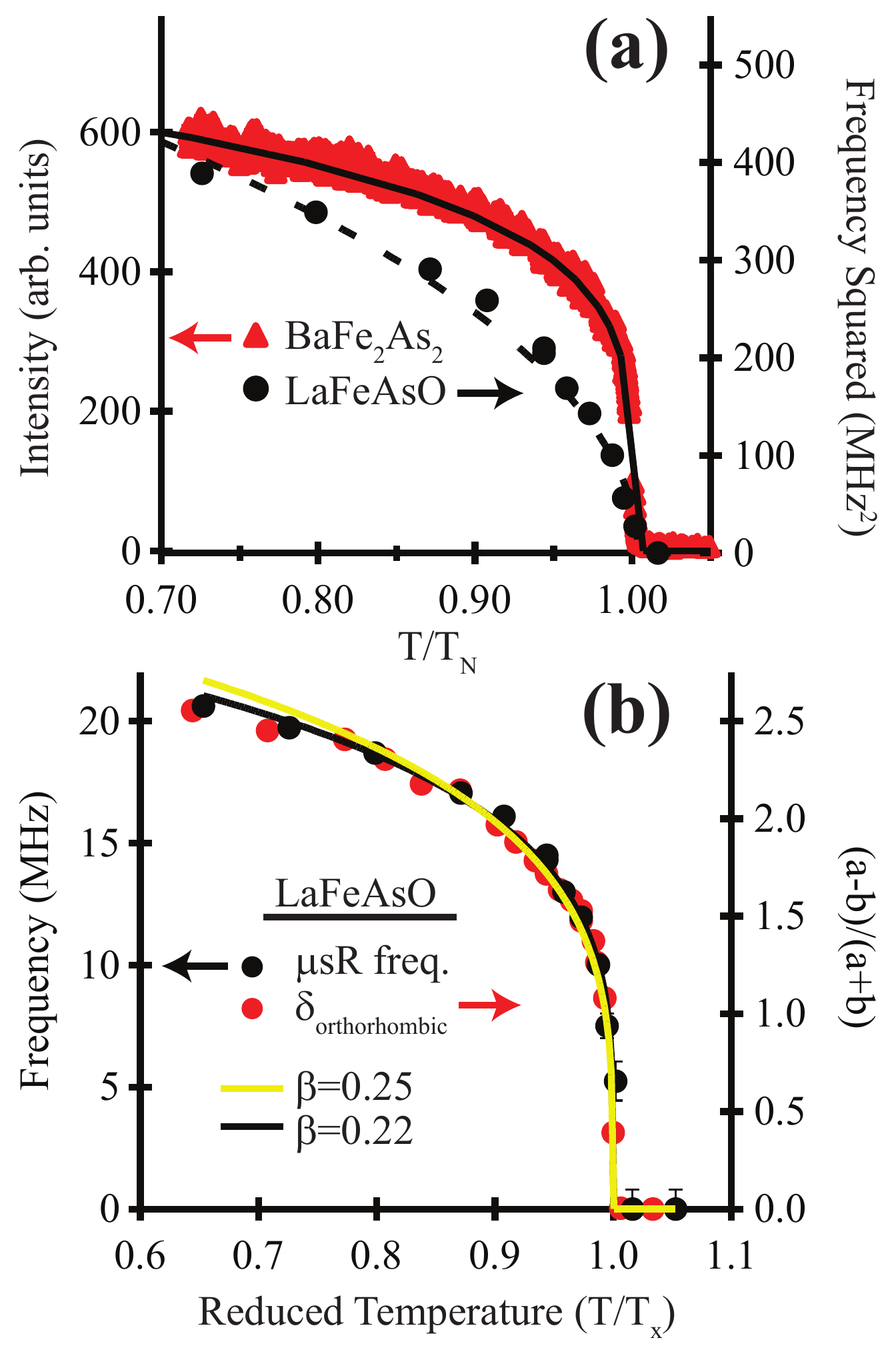}
\caption{(a) Plot showing the comparison of M$^{2}$($\frac{T}{T_N}$) for BaFe$_2$As$_2$ and LaFeAsO.  The solid black line denotes the Gaussian convolved power law fit from Fig. 2 (b), and the dashed line shows a power law fit with $\beta=0.25$ as described in the text.  (b) Overplot of the magnetic order parameter (data taken from Ref. 22) and the structural order parameter (data taken from Ref. 23) for the LaFeAsO system.}
\label{fig:Fig3}
\end{figure}

\section{Magnetic and Structural Order Parameters in the Iron Arsenides} 

Having established that the magnetic phase transition in Ba-122 mirrors that of a known 2D-Heisenberg-Ising magnet, it is informative now to compare the magnetic order parameters between different classes of iron arsenides.  In Fig. 3 (a), measurements of magnetic order parameters squared are plotted for both Ba-122 and LaFeAsO, where an immediate difference between the temperature dependencies of the order parameters is apparent. The data for LaFeAsO are taken from Ref. 22 \nocite{maeter} which also reports a clear commonality among the measured magnetic order parameters of the REFeAsO systems.  Therefore from Fig. 3 (a) it is clear that there is a fundamental difference in the phase transition behavior of the Ba-122 and the 1111 class of iron arsenides where the order parameter in LaFeAsO fit between $0.6<\frac{T}{T_N}<1.05$ to a simple power law gives a $\beta=0.22\pm0.06$. The dashed line in Fig. 3 (a) shows a simple power law function with a fixed $\beta=0.25$ overplotted with the LaFeAsO data.  Unfortunately, due to the lack of detailed data close to the phase transition in LaFeAsO, the critical exponent for the magnetic order parameter cannot be determined precisely; however from Fig. 3 (a) it is apparent that a simple $\beta=0.25$ provides a reasonable approximation to the magnetic phase behavior in this system. This renders a critical exponent $\beta$ in the monolayer iron arsenides roughly two times larger than the $\beta=0.125$ reported to model the phase behavior in Ba-122.\cite{wilson} Given that both systems possess the same intrinsic spin structures and magnetic ions, this sharp distinction is surprising and, we believe, of fundamental importance. 

In noting that the strongly coupled structural and magnetic order parameters in Ba-122 are known to show an identical temperature evolution,\cite{wilson} it is clearly important to examine how the structural phase transition is modified in LaFeAsO where T$_S$ and T$_N$ are largely decoupled.  An overlay of both the magnetic order parameter data again taken from Ref. 22 \nocite{maeter} and the structural order parameter data extracted from Ref. 23 \nocite{quan} is plotted in Fig. 3 (b).  From this figure, it is immediately evident that, despite the splitting of T$_S$ and T$_N$ by over more than $14$K, the phase behaviors of both the magnetic and structural order parameters in LaFeAsO track one another quite precisely. The yellow solid line in Fig. 3(b) shows the results of a power law fit with $\beta=0.25$ overlaid with both sets of data while the black line shows the results from the freely refined fit of the magnetic order parameter in LaFeAsO (discussed previously).  This simple, empirical, comparison suggests that the magnetostructural coupling observed in Ba-122 and other undoped, bilayer pnictides is also reflected in the monolayer pnictides where T$_S$ and T$_N$ are no longer coincident.  The primary effect of tuning away from the T$_N$=T$_S$ critical point appears to be a renormalization of the critical exponents modeling both the structural and magnetic order parameters which are altered in the same fashion.  Despite the significant difference in $\beta$ values between the 122 and 1111 compounds, the structural and magnetic phase behaviors mirror one another identically in both systems, thus strongly suggesting that this is a universal feature within the parent phases of the iron arsenides.              
 
In considering the immediate contrasts in the magnetic phases of the 122 and 1111 compounds which may lead to the large differences in their magnetic and structural phase transition behaviors, the primary distinction is that the strongly coupled, concomitant, magnetostructural phase transitions in the 122 series splits in temperature within the 1111 compounds thereby potentially altering the critical spin behavior. In order to investigate this further, we examined data from the literature reporting the magnetic order parameters in lightly Ni- and Co-doped Ba-122 systems\cite{lelandBa122Ni, mcqueeneyBa122Co, christiansonBa122Co}.  Within these initial reports, it was observed that doping small amounts of electrons or holes splits the magnetic and structural phase transitions in Ba-122, and from the reported data it can be clearly seen that the observed order parameter is dramatically modified from the parent system's behavior.  In both cases, a simple power law with an exponent of $\beta\approx0.3$ was reported to model the magnetic phase transitions close to T$_N$\cite{mcqueeneyBa122Co, lelandBa122Ni}.  In order to provide a more explicit comparison between these various systems, Fig. 4 overplots the magnetic order parameters of both of these doped Ba-122 systems along with the magnetic order parameters of several different iron arsenide parent systems extracted from the literature\cite{zhaoelastic, mcqueeneyBa122Co, lelandBa122Ni, shiliangNaFeAs, wilson, christiansonBa122Co} .  Systems whose phase transitions are closely modeled by a $\beta=0.125$ were normalized together at $T/T_N=0.9\approx0.56$M$^{2}$($\frac{T}{T_N}=0$), and systems whose phase transitions were approximated by $\beta\approx0.25$ were normalized together at $T/T_N=0.68\approx0.56$ M$^{2}$($\frac{T}{T_N}=0$).  For clarity, the order parameters of each class of systems were set apart by cross-normalizing the average values of the order parameters of the $\beta\approx0.25$ class and the $\beta=0.125$ class at $75\%$ of their saturated values.  In determining which data to include in this comparison, we attempted to select the highest quality data in the literature using single crystal measurements when possible.  As Sn-flux grown crystals exhibit Sn incorporation which strongly renormalizes magnetic properties in certain 122 pnictides\cite{su}, we utilized data only from studies performed on crystals grown via FeAs or self-flux methods.

\begin{figure}
		\includegraphics[scale=0.4]{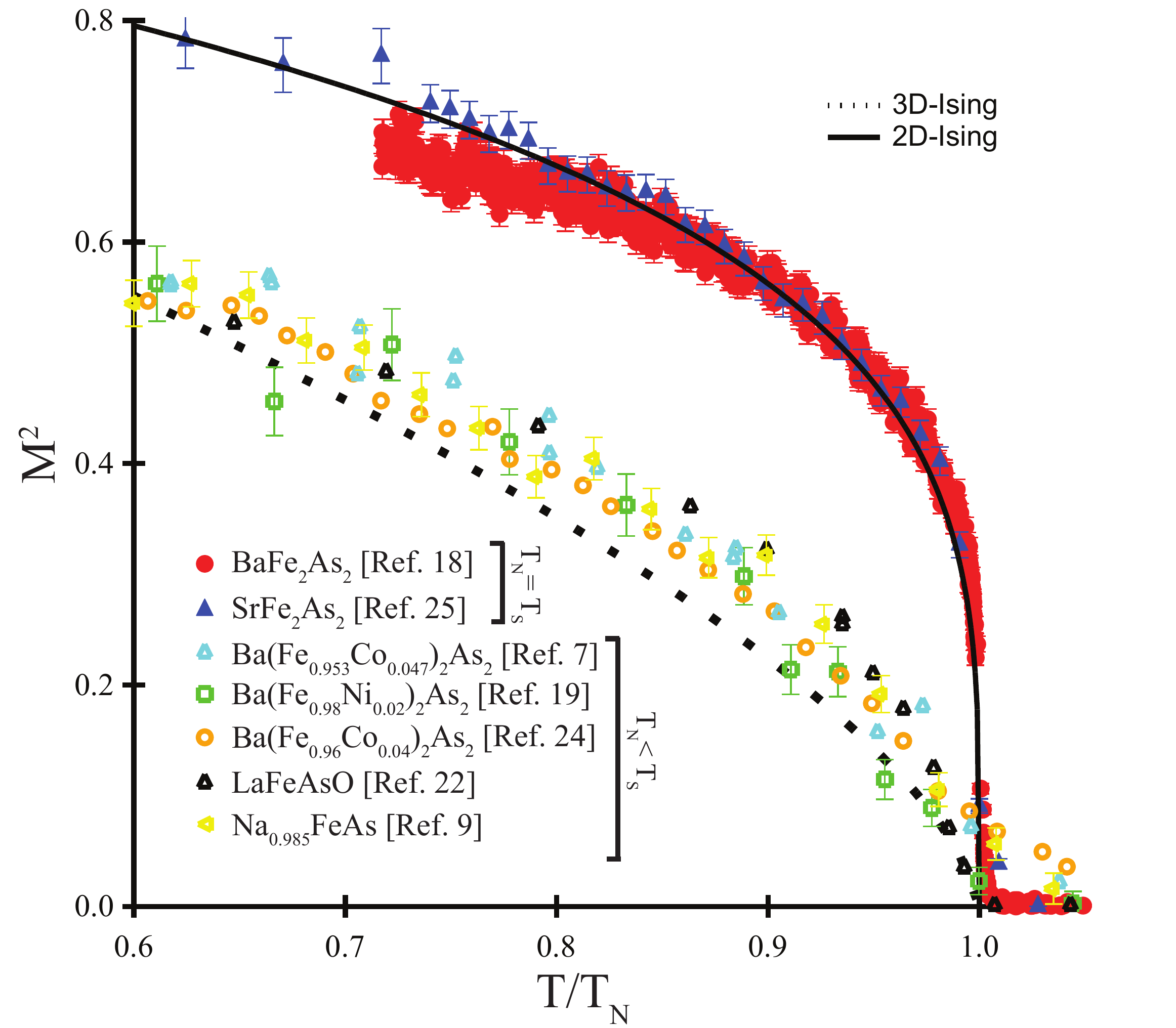}
		\caption{Plot showing the comparison of M$^{2}$($\frac{T}{T_N}$) for systems in the 122, 1111, and 111 classes of iron arsenide superconductors with the normalization procedure between different systems described in the text.  The solid black line  and dashed black line denote the ideal 2D-Ising and 3D-Ising values for M$^{2}$($\frac{T}{T_N}$) respectively.  The collapse onto two distinct curves for systems where T$_N$=T$_S$ and for systems where T$_N<$T$_S$ is shown.}
	\label{fig:Fig4}
\end{figure}

Looking at Fig. 4, it is immediately clear that there are two distinct classes of transitions amongst the data plotted.  Both 122 parent systems, SrFe$_2$As$_2$ and BaFe$_2$As$_2$, collapse onto a single curve that is modeled well by the $\beta=0.125$ 2D-Ising order parameter. Data from doped 122 systems and from 1111 and 111 iron arsenide compounds in Fig. 4 reveal that, in systems where the structural and magnetic phase transitions separate in temperature, the resulting magnetic phase transition collapses onto a separate universal curve whose critical exponent is roughly double that of the $T_N=$T$_S$ materials.  Within this second class of magnetic phase transitions, fits to the data between $0.6\leq\frac{T}{T_N}\leq1.1$ led to an average exponent of $\beta=0.27\pm0.03$ demonstrating that, once the structural phase transition decouples from the onset of magnetic order, the universality class of the magnetic phase transition is clearly altered.  Currently however, the quality of data from these systems with broader, decoupled magnetic phase transitions precludes the determination of their precise critical exponents; however we have plotted in Fig. 4 one possibility of a crossover to a 3D-Ising order parameter with $\beta=0.325$. A 3D-Heisenberg transition with $\beta=0.365$ (not shown) may also fall within the error of the data suggesting in both cases an increased dimensionality to the magnetic phase transition once decoupled from T$_S$.

\section{Discussion}
Several theories have been put forth examining the relationship between the higher temperature structural transition and the lower temperature magnetic phase transition in the 1111 series of pnicides.  Work by Xu et al. considered the presence of a finite temperature 2D Ising order which can be envisioned as an electronic, nematic phase that forms prior to the onset of the spin density wave (SDW) order.\cite{xu}  Additionally, recent work by Chen et al. has considered the presence of an orbital ordering parameter also possessing 2D-Ising symmetry stemming from a simple spin-orbital Hamiltonian.\cite{chentheory}  SDW order in these models freezes out due to the finite J$_c$ coupling and is simply a reflection of the eventual onset of three-dimensional exchange within the system.  However, to the best of our knowledge, no theoretical work exploring the expected spin behavior at the effective critical point in the 122 pnictides where both phase transitions are coincident in temperature has been performed.  

Experimental work on BaFe$_2$As$_2$ has reported a reduction in the anisotropy gap of the spin excitations upon Ni-doping onto the Fe-site, potentially indicating an enhanced two dimensionality induced upon carrier doping.\cite{lelandBa122Ni}  This seems to run counter to our current observation of a transition from a regime in which the magnetic phase transition behavior is consistent with that of a two dimensional magnetic system in the BaFe$_2$As$_2$ system to a regime in doped samples whose larger critical exponents (ie. $\beta\approx0.27$) suggest a new universality class with a more three dimensional character.  Recent photoemission experiments however have shown an abrupt change in the dimensionality of Co-doped Ba-122 where the two dimensional electronic states of the parent Ba-122 system are tuned toward a more three dimensional dispersion as Co is doped into the system.\cite{Thirupathaiah}  Separate neutron studies of BaFe$_2$As$_2$ have also reported largely two dimensional spin fluctuations above T$_S$\cite{matan} consistent with the picture of a two dimensional magnetic order parameter close to T$_N$ and qualitatively in agreement with the largely two dimensional Fermi surface reported in recent photoemission measurements on BaFe$_2$As$_2$\cite{Thirupathaiah} and CaFe$_2$As$_2$.\cite{liuCaFe2As2}  To-date however, there has been no comprehensive experimental investigation into the detailed evolution of the dimensionality of spin fluctuations in a 122 system as it is tuned from the quasi-2D high temperature regime above T$_S$ to the anisotropic, three dimensional magnetic and electronic phase far below T$_N$.  Future work exploring this along with further theoretical efforts detailing the coupled phase transitions in the pnictides are needed in order to understand the seemingly universal structural and magnetic phase behaviors within these materials.              

\section{Conclusion}
In summary, we have presented an empirical analysis of the magnetic order parameters in the primary classes of the iron arsenide superconductors.  Within the undoped, bilayer 122 class of iron pnictides where T$_N$=T$_S$, both the magnetic and structural phase transitions are well modeled by a 2D-Ising order parameter; however, upon doping, the magnetic and structural phase transitions alter to exhibit a critical $\beta\approx0.27$.  The likely cause of this crossover in the phase behavior is the decoupling of the magnetic and structural transitions where alternate classes of iron pnictides with T$_N<$T$_S$ collapse onto the same seemingly universal curve.  Two distinct behaviors therefore appear in the magnetic and structural phase transitions of the iron arsenides:  The first of these occurs when the structural and magnetic phase transitions coincide at a multicritical point, and the second emerges once these two transitions no longer coincide rendering strongly renormalized critical exponents for both.  The critical behavior of the iron pnictides therefore seems to transition from one which parallels a known 2D magnet when T$_N$=T$_S$ to one consistent with a more three dimensional character upon tuning away from the critical point.  This observation currently contradicts several existing arguments for more two dimensional behavior reported in the Ba-122 system upon introducing charge carriers, thus suggesting that new theoretical insight is needed.  Our analysis demonstrates that further theoretical and experimental work exploring the phase behavior in the critical magnetism of the iron pnictides have the possibility of resolving exciting new physics relevant to the fundamental symmetries in the magnetism of these compounds.

\begin{acknowledgments}
We would like to thank P. Dai, S. Li, L. Harriger, J. Zhao, and H. Maeter for providing access to their raw data for our analysis.  This work was supported by the Director, Office of Science,  Office of Basic Energy Sciences, U.S. Department of Energy, under Contract No. DE-AC02-05CH11231 and Office of Basic Energy Sciences US DOE under Contract No.DE-AC03-76SF008.

\end{acknowledgments}

\end{document}